\documentclass[twocolumn,pre,superscriptaddress,floatfix,longbibliography]{revtex4-1}

\usepackage{graphics}
\usepackage{graphicx}
\usepackage{amsmath}
\usepackage{amssymb,xcolor}
\usepackage{subcaption} % Additional package

\newcommand{\textin}[1]{\mbox{\scriptsize{#1}}}

\definecolor{grisclair}{rgb}{0.6,0.6,0.6}

\newcommand{\beq}{\begin{equation}}
\newcommand{\ee}{\end{equation}}

\begin{document}

\title{Stability analysis of the flow in a coflowing device}
\author{M. Rubio}
\address{Depto.\ de Ingenier\'{\i}a Energ\'etica y Fluidomec\'anica and\\
Instituto de las Tecnolog\'{\i}as Avanzadas de la Producci\'on (ITAP),\\
Universidad de Valladolid, E-47003 Valladolid, Spain}
\author{ S. Rodríguez-Aparicio}
\address{Depto.\ de Ingenier\'{\i}a Mec\'anica, Energ\'etica y de los Materiales and\\ 
Instituto de Computaci\'on Cient\'{\i}fica Avanzada (ICCAEx),\\
Universidad de Extremadura, E-06006 Badajoz, Spain}
\author{M. G. Cabezas}
\address{Depto.\ de Ingenier\'{\i}a Mec\'anica, Energ\'etica y de los Materiales and\\ 
Instituto de Computaci\'on Cient\'{\i}fica Avanzada (ICCAEx),\\
Universidad de Extremadura, E-06006 Badajoz, Spain}
\author{J. M. Montanero}
\address{Depto.\ de Ingenier\'{\i}a Mec\'anica, Energ\'etica y de los Materiales and\\ 
Instituto de Computaci\'on Cient\'{\i}fica Avanzada (ICCAEx),\\
Universidad de Extremadura, E-06006 Badajoz, Spain}
\author{M. A. Herrada}
\address{E.T.S.I., Depto.\ de Ingenier\'{\i}a Aeroespacial y Mec\'anica de Fluidos, Universidad de Sevilla, Camino de los Descubrimientos s/n 41092, Spain}

\begin{abstract}
We analyze the stability of the coflow configuration. The experiments and the global stability analysis show that the emitted jet always destabilizes before the tapering conical meniscus. This implies that the parameter conditions at which polydisperse dripping arises cannot be determined from the linear stability analysis of the steady jetting mode. Transient simulations show that the linear superposition of decaying eigenmodes triggered by an initial perturbation can lead to the jet breakup. The breakup process significantly depends on the initial perturbation. These results question the validity of the linear stability analysis as applied to the coflowing and other similar configurations. 
\end{abstract}

\maketitle

\section{Introduction}
\label{sec1}

% Capillary and absolute instabilities
Capillary jets are intrinsically unstable due to the growth of capillary waves. Convectively unstable jets sweep all the perturbations downstream, preserving both a stable liquid source and a steady filament, long compared to its diameter. In this case, quasi-monodisperse droplets arise at the end of the jet due to the Rayleigh-Plateau capillary instability \citep{R92b}. This is the so-called jetting mode commonly used to produce droplets continuously. Conversely, absolute instability allows perturbations to travel upstream or remain in the liquid source while growing \citep{LG86a,HM90a}, preventing the jet formation and producing droplets right behind the liquid source. This is usually referred to as the dripping mode. Dripping realizations involving the oscillation of the liquid source are undesired because they typically produce polydisperse droplets. 

% Global modes. Global stability.
Global linear stability analysis is the preferred tool to determine the stability of flows in multiple fields, including multiphase capillary flows in microfluidics. Global linear eigenmodes are motion patterns in which the entire system moves harmonically with the same (complex) frequency and a fixed phase relation \citep{T11}. They are calculated as the eigenfunctions of the linearized Navier-Stokes operator as applied to a steady configuration (base flow). Suppose the spectrum of eigenvalues is in the stable complex half-plane. In that case, any initial small-amplitude perturbation will decay exponentially with time for $t\to \infty$, and the base flow is {\em asymptotically} (for large $t$) stable. If at least one of the eigenvalues is in the unstable complex half-plane, the perturbation will grow exponentially (as long as the linear approximation applies), and the base flow is unstable.

% Global stability. Success and failure
Global linear stability analysis has successfully described the absolute instability (the jetting-to-dripping transition) of several capillary flows, such as gravitational jets \citep{SB05a}, coflow \citep{GSC14}, gaseous \citep{CHGM17,RRCHGM21} and liquid-liquid \citep{CRRHM21} flow focusing, confined selective withdrawal \citep{LCMH22}, and electrospray \citep{PRHGM18}. This method is also expected to predict the convective instability of the jetting regime (the jet breakup), provided that the analyzed fluid domain encloses the region where the jet breaks up. However, previous attempts have failed to determine, for instance, the jet breakup length \citep{GCHWKGLCHMB21}. Several explanations have been proposed to justify this failure \citep{U16,GCHWKGLCHMB21,M24}. 

% The short-term response
When an asymptotically stable base flow is perturbed, the superposition of the decaying eigenmodes triggered by the initial perturbation can lead to the short-term growth of the perturbation energy, making the flow unstable \citep{S07}. This phenomenon may explain the appearance of convective instabilities not predicted by the global stability analysis. The non-modal growth of perturbations allows one to describe convective instabilities in problems with inflow and outflow conditions, including gravitational jets \citep{LCC02}, gaseous flow focusing \citep{CHGM17}, and ballistic capillary jets \citep{HMH21}.

% Coflow
Hydrodynamic forces can control the pinching of fluid-liquid interfaces and produce the microdripping and microjetting modes of tip streaming \citep{M24}. The coflowing device is probably the simplest hydrodynamic configuration to produce tip streaming. In a coflowing device, two immiscible fluids are coaxially injected across cylindrical tubes with radii $R_i$ and $R_o$. Either jets or droplets are emitted from the tip of a fluid cone stretched by the outer viscous force. We can distinguish between coflowing realizations with $R=R_o/R_i$ of order unity \citep{SB06,GCUA07,GCA08,GSC14} and those with $R\gg 1$ \citep{CFW04,UFGW08}. Their behaviors are substantially different owing to the stabilizing role of confinement \citep{CHM19b} and the influence of this factor on the nonlinear phase of the jet breakup. Using a slender-body approximation and neglecting inertia effects, \citet{GSC14} described the transition from monodisperse to polydisperse dripping from the global stability analysis of an unconfined system ($R\ll 1$) in the limit of a very small inner-to-outer viscosity ratio.

% This paper
The present paper analyzes the stability of the coflow configuration numerically and experimentally. Our theoretical model has no limitations regarding the Reynolds number and the viscosity ratio. We find that the global stability analysis cannot predict the jetting-to-dripping transition. The experiments show that the interface oscillations caused by the convective instability progressively displace upstream as the dispersed-phase flow rate decreases, eventually leading to the dripping mode. Consequently, the marginally stable flow at the jetting-to-dripping transition is an oscillatory state in which droplets are produced at the tip of the steady fluid cone. This state is very different from the steady jetting base flow considered in the global stability analysis, which explains the failure of this method. 

The major contribution of this work is the analysis of the droplet production in the jetting regime. The transient numerical simulations allow us to explain this in terms of the non-exponential short-term response of the base flow to an initial perturbation. The interference of decaying eigenmodes produces the jet breakup before those modes are damped. 

% Organization 
The paper is organized as follows. We formulate the problem in Sec.\ \ref{sec2}. The governing equations and some details of the numerical procedure are shown in Sec.\ \ref{sec3}. The experimental method is described in the Supplemental Material. We analyze the jetting-to-dripping transition in Sec.\ \ref{seca1}. The flow convective instability is studied in Secs.\ \ref{seca2} and \ref{seca3}. The paper closes with some concluding remarks in Sec.\ \ref{sec5}. 

\section{Formulation of the problem}
\label{sec2}

% Parameters
In the coflowing configuration, a liquid of density $\rho_i$ and viscosity $\mu_i$ is injected across a cylindrical inner capillary of radius $R_i$ at a constant flow rate $Q_i$. This liquid flows surrounded by an outer stream of density $\rho_o$ and viscosity $\mu_o$. The outer current is injected at a constant flow rate $Q_o$ across a cylindrical capillary of radius $R_o$ concentric with the inner one. The two fluids are immiscible, and the surface tension is $\gamma$. For sufficiently small values of $Q_i$, the inner liquid accelerates within a fluid meniscus, tapering a jet much thinner than the inner feeding capillary.

% Characteristic quantities. Without surfactant
The problem can be formulated in terms of the ratio $R=R_o/R_i$, the density and viscosity ratios, $\rho=\rho_o/\rho_i$ and $\mu=\mu_o/\mu_i$, the Ohnesorge and capillary numbers,
\begin{equation}
\text{Oh}_i=\frac{\mu_i}{(\rho_i \gamma R_i)^{1/2}} \quad \text{and} \quad Ca=\frac{\mu_o U}{\gamma},
\end{equation}
and the flow rate ratio $Q=Q_i/Q_o$. Here, $U=Q_o/[\pi (R_o^2-R_i^2)]$ is the mean outer velocity upstream. 

% Control parameters
In a typical experimental run, the microfluidic device and the fluids are fixed. For this reason,  $\{R$, $\rho$, $\mu$, Oh$_i\}$ take constant values while $Ca$ and $Q$ are the control parameters. To determine the stability limit, one selects the outer flow rate $Q_o$ and progressively decreases the inner flow rate $Q_i$ until its minimum value $Q_{i\textin{min}}$ for stable microjetting is reached. The experiment is usually repeated for several values of $Q_o$. Therefore, the goal is to determine the dimensionless minimum flow rate $Q_{\textin{min}}=Q_{i\textin{min}}/Q_o$ as a function of $Ca$. 

\section{Governing equations and numerical method}
\label{sec3}

% Bulk equations
The inner capillary radius $R_i$, the visco-capillary velocity $v_{\gamma\mu}=\gamma/\mu_i$, and the capillary pressure $\gamma/R_i$ are the characteristic quantities. We integrate the dimensionless Navier-Stokes equations for the axisymmetric velocity $\mathbf{v}^{(k)}(r,z;t)$ and pressure $p^{(k)}(r,z;t)$ fields:
\begin{equation}
\left(ru^{(k)}\right)_r+rw^{(k)}_z=0, \label{basic1}
\end{equation}
\begin{eqnarray}
&&\rho^{\delta_{ko}}\, \left(\text{Oh}_i\right)^{-2}\left(\frac{\partial u^{(k)}}{\partial t} + u^{(k)} u^{(k)}_r+ w^{(k)} u^{(k)}_z\right)\nonumber\\
&&=-p^{(k)}_r+\mu^{\delta_{ko}}
\left(u^{(k)}_{rr}+(u^{(k)}/r)_r+u^{(k)}_{zz}\right),\nonumber\\ \label{basic2}
\end{eqnarray}
\begin{eqnarray}
&&\rho^{\delta_{ko}}\, \left(\text{Oh}_i\right)^{-2}\left(\frac{\partial w^{(k)}}{\partial t} + u^{(k)} w^{(k)}_r+ w^{(k)}w^{(k)}_z\right)\nonumber\\
&&=-p^{(k)}_z+\mu^{\delta_{ko}\,} \left(w^{(k)}_{rr}+w^{(k)}_r/r+w^{(k)}_{zz}\right),\nonumber \\
\label{basic3}
\end{eqnarray}
where $t$ is the time, $r$ ($z$) is the radial (axial) coordinate, $u^{(k)}$ ($w^{(k)}$) is the radial (axial) velocity component, and $\delta_{ko}$ is the Kronecker delta. In the above equations and henceforth, the superscripts $k=i$ and $o$ refer to the inner and outer phases, respectively. In addition, the subscripts $r$ and $z$ denote the partial derivatives with respect to the corresponding coordinates. The action of the gravitational field has been neglected due to the smallness of the fluid configuration.

\begin{figure}
\begin{center}
\includegraphics[width=8cm]{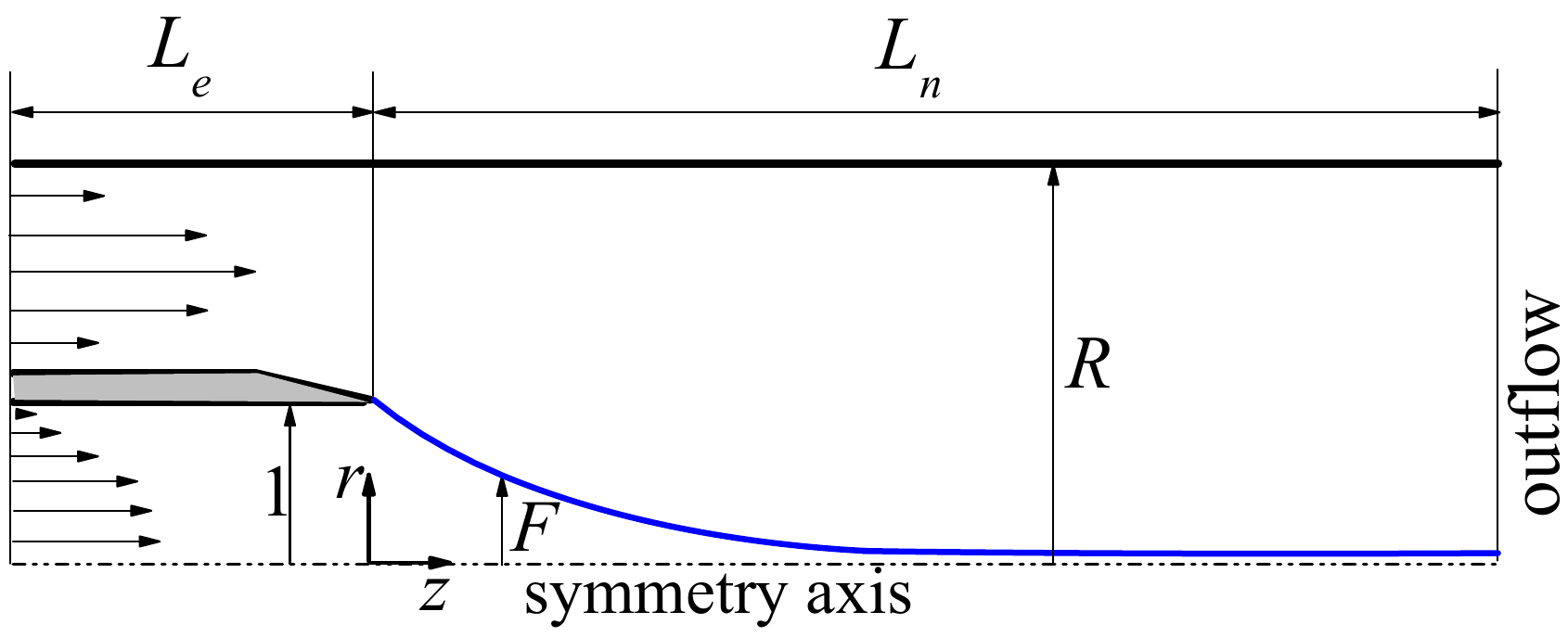}
\end{center}
\caption{Sketch of the numerical domain.}
\label{numer1}
\end{figure}

% Interface boundary conditions
The velocity field is continuous at the interface, i.e., 
\begin{equation}
\mathbf{v}^{(i)}=\mathbf{v}^{(o)}
\end{equation}
at $r=F(z,t)$, where $F(z,t)$ is the distance of an interface element to the symmetry axis (Fig.\ \ref{numer1}). The kinematic compatibility at the interface yields
\begin{eqnarray}
\frac{\partial F}{\partial t}+F_z w^{(i)}-u^{(i)}=0.\label{int1}
\end{eqnarray}
The equilibrium of normal stresses on that surface leads to
\begin{equation}
\label{n}
-p^{(i)}+\tau_n^{(i)}=\gamma\kappa-p^{(o)}+\tau_n^{(o)},
\end{equation}
where
\begin{equation}
\label{cur}
\kappa=\frac{FF_{zz}-1-F_z^{2}}{F(1+F_z^{2})^{3/2}}
\end{equation}
is the local mean curvature, and 
\begin{equation}
\tau_n^{(i)}=\frac{2[u^{(i)}_r-F_z(w^{(i)}_r+u^{(i)}_z)+F_z^{2}w^{(i)}_z]}{1+F_z^{2}},
\end{equation}
\begin{equation}
\tau_n^{(o)}=\frac{2\mu[u^{(o)}_r-F_z(w^{(o)}_r+u^{(o)}_z)+F_z^{2}w^{(o)}_z]}{1+F_z^{2}}
\end{equation}
are the inner and outer normal viscous stresses, respectively. The equilibrium of tangential stresses yields
\begin{eqnarray}
\label{t}
\tau_t^{(i)}=\tau_t^{(o)}, \label{int2}
\end{eqnarray}
where $\tau_t^{(i)}$ is the inner tangential viscous stress and $\tau_t^{(o)}$ is the outer tangential viscous stress given by the expressions
\begin{equation}
\label{t1}
\tau_t^{(i)}=(1-F_z^{2})(w^{(i)}_r+u^{(i)}_z)+2F_z(u^{(i)}_r-w^{(i)}_z),
\end{equation}
\begin{equation}
\label{t3}
\tau_t^{(o)}=\mu[(1-F_z^{2})(w^{(o)}_r+u^{(o)}_z)+2F_z(u^{(o)}_r-w^{(o)}_z)].
\end{equation}

% Boundary conditions
The hydrodynamic equations are integrated within the numerical domain sketched in Fig.\ \ref{numer1}. We prescribe parabolic velocity distributions at the inlet sections of the inner and outer capillaries ($z=-L_e$). In all simulations, $L_e=15$. We consider the no-slip boundary condition at the solid walls and the outflow conditions 
\begin{equation}
\label{out}
u^{(k)}_z=w^{(k)}_z=F_z=0    
\end{equation}
at the right-hand end of the computational domain. The anchorage condition of the triple contact line, $F=1$, is imposed at the edge of the feeding capillary. We verified the results are essentially the same when the Hagen-Poiseuille velocity profile is prescribed instead of the outflow boundary condition (\ref{out}). Finally, the standard regularity conditions $u^{(i)}=w^{(i)}_r=0$ are prescribed at the symmetry axis. 

% Global modes
In the asymptotic global stability analysis, we assume the existence of a perfectly steady base flow within the numerical domain (Fig.\ \ref{numer1}) and determine whether small amplitude perturbations destabilize that flow. We consider the following temporal dependence for the perturbations: 
\begin{eqnarray}
U(r,z;t)&=&U_0(r,z)+\delta U(r,z)\, e^{-i\omega t}+\text{c.c.} \quad (|\delta U|\ll U_0), \nonumber\\
F(z;t)&=&F_0(z)+\delta F(z)\, e^{-i\omega t}+\text{c.c.}\quad (|\delta F|\ll F_0),
\label{perr}
\end{eqnarray}
where $U(r,z;t)$ represents the velocity and pressure fields, while $U_0(r,z)$ and $\delta U(r,z)$ stand for the base flow (steady) solution and the spatial dependence of the eigenmode, respectively. In addition, $F_0(z)$ represents the base flow solution for $F(z;t)$, while $\delta F(z)$ is the corresponding perturbation amplitude. The perturbation evolves according to the eigenfrequency $\omega=\omega_r+i\omega_i$, where $\omega_r$ and $\omega_i$ are the oscillation frequency and growth rate, respectively. The flow asymptotic linear stability \citep{T11} is determined by the eigenfrequency $\omega^*=\omega_r^*+i\omega_i^*$ of the dominant mode (that with the largest growth rate). Realizations with $\omega_i^*<0$, $\omega_i^*=0$, and $\omega_i^*>0$ correspond to stable, marginally stable, and unstable flows, respectively.

% DNS
We conduct transient (direct) numerical simulations of the linearized equations to study the base flow response to an initial perturbation. The perturbation consists of the deformation of the free surface (the velocity and pressure fields are not perturbed) at $t=0$ given by the function
\begin{equation}
\label{def}
F(z;0)-F_0(z)=\beta\ e^{-\alpha |z-z_0|}\; ,
\end{equation}
where $\beta$ indicates the maximum deformation, $z_0$ and $\alpha$ are the location and decay rate, respectively. All the simulations were conducted for $\beta=10^{-3}$ and $\alpha=10$. 

The perturbation (\ref{def}) triggers a train of capillary waves propagating downstream. For sufficiently large $t$, the contribution of subdominant eigenmodes is expected to become negligible, and the dominant mode is supposed to govern the system's linear dynamics. In this case, the sign of the dominant mode growth rate is assumed to determine flow stability; i.e., the perturbations are damped for $\omega_i^*<0$ and grow until producing the interface breakup for $\omega_i^*>0$. 

% The comparison
It must be noted that nonlinear terms were not considered in the transient numerical simulations. Therefore, the hydrodynamic equations, the numerical domain, and the grid used in the calculations were the same in the linear stability analysis and the transient simulations. While the linear stability analysis determines the evolution of the eigenmodes separately, the transient simulations consider the superposition of those modes, solving the same algebraic equations in both cases. The full hydrodynamic equations were integrated to analyze the nonlinear effects at the end of Sec.\ \ref{sec4}.

% Numerical method.
The governing equations are integrated with a variant of the numerical method proposed by \citet{HM16a} (see also Ref.\ \citep{JAM}). We use a boundary-fitted spectral method \citep{HM16a} to solve the theoretical model described above. The inner and outer fluid domains are mapped onto two quadrangular domains through non-singular mapping. A quasi-elliptic transformation \citep{DT03} is applied in the outer bath. All the derivatives appearing in the governing equations are expressed in terms of the spatial coordinates resulting from the mapping. These equations are discretized in the mapped radial direction $\eta$ with Chebyshev spectral collocation points \citep{KMA89}. Specifically, we use $n_{\eta}^{(i)}=21$ and $n_{\eta}^{(o)}=85$ in the inner and outer domains, respectively. We use fourth-order finite differences with $n_{\xi}=80$ ·  $(L_{e}+L_{n})$ equally spaced points to discretize the mapped axial direction $\xi$. The MATLAB {\sc eigs} function is applied to find the eigenfrequencies around a reference value $\omega_0$. This process is repeated for several values of $\omega_0$. In the transient numerical simulations, second-order backward finite differences are used to discretize the time domain. The time step $\Delta t=0.1$ is constant. 

% Grid sensitivity analysis
We verified that the discretization errors significantly affected neither the eigenvalues nor the transient simulations. For instance, the growth rate $\omega_i$ of the critical mode analyzed in Fig.\ 6 ($L_n=20$) differed in less than 2\% when the number of grid points increased from 254612 to 453852. The transient simulations were practically insensitive to significant variations of $\Delta t$. The Supplemental material shows the grid used in the simulations.

\section{Results}
\label{sec4}

We devote Sec.\ \ref{seca1} to analyzing the transition from jetting to polydisperse dripping. Sections \ref{seca2} and \ref{seca3} study the convective instability in the jetting regime. All the simulations and experiments were conducted for $R=6.67$, $\rho=0.962$, $\mu=36.1$, and Oh$_i=0.0264$. 

\subsection{The jetting-to-dripping transition}
\label{seca1}

% Tip streaming. Qmin
In all tip streaming configurations, for fixed values of the rest of the governing parameters, there is a minimum value $Q_{\textin{min}}$ of the flow rate ratio $Q$ below which the tapering meniscus oscillates and emits polydisperse droplets \citep{M24}. This transition from jetting to polydisperse dripping results from the so-called flow's absolute instability, in which perturbations cannot be convected downstream and destabilize the meniscus. In most tip streaming configurations, there is correspondence between what occurs in the experiment and what is assumed in the linear stability. In both cases, for $Q>Q_{\textin{min}}$, there is a perfectly steady base flow characterized by a long jet in which the perturbations are either damped or swept downstream beyond the fluid domain, while those perturbations destabilize the tapering meniscus for $Q<Q_{\textin{min}}$. The linear stability analysis allows one to calculate the amplitude $\delta F(z)$ of the interface deformation corresponding to the critical eigenmode responsible for this instability transition. This amplitude shows that the tapering meniscus oscillates for $Q< Q_{\textin{min}}$, as occurs in the experiments \citep{M24}.  

\begin{figure*}
\begin{center}
\includegraphics[width=12cm]{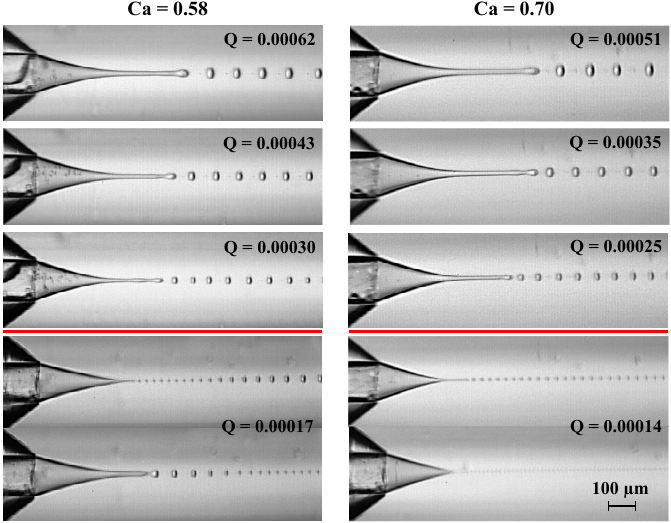}
\end{center}
\caption{Experimental images for different values of $Ca$ and $Q$. The red horizontal line indicates the transition to the dripping mode. The two lower images show the tapering meniscus oscillation for the same value of $Q$.}
\label{modes}
\end{figure*}

% Coflow. Experiment
The experiments of the coflow configuration show a behavior significantly different from that described above. The steady jetting flow mentioned above is not established for $Q\gtrsim Q_{\textin{min}}$. Conversely, the flow adopts a periodic state that suffers from convective instability. In this state, the tapering meniscus remains steady while quasi-monodisperse droplets are produced from the breakage of a stretched liquid filament close to the meniscus tip (Fig.\ \ref{modes}). The filament length decreases, and its shape deviates from the cylindrical one as $Q$ approaches $Q_{\textin{min}}$. For flow rate ratios right above $Q_{\textin{min}}$, there is no distinction between the stretched tapering meniscus and the liquid filament. For $Q< Q_{\textin{min}}$ (Fig.\ \ref{modes}, below the red horizontal line), the stretched meniscus oscillates (expands and retracts), giving rise to polydisperse droplets (dripping mode). We conclude that the conical meniscus remains stable even for flow rate ratios leading to jet instability. \citet{GSC14} have shown that the meniscus stability is caused by its axial curvature.

\begin{figure}
\begin{center}
\includegraphics[width=0.9\linewidth]{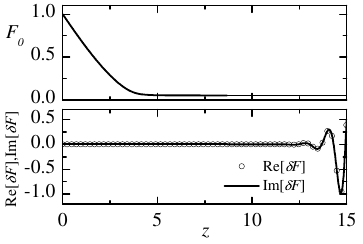}
\end{center}
\caption{Real (a) and imaginary (b) part of $\delta F$ for $Ca=0.58$ and $Q=0.000125$.}
\label{amplitude2}
\end{figure}

% Coflow. Numerics
In the global stability analysis of the coflow configuration, we followed the same procedure as in the rest of the tip streaming realizations \citep{CHGM17,RRCHGM21,CRRHM21,LCMH22,PRHGM18}. We assumed the existence of a perfectly steady base flow and wondered whether perturbations destabilize that flow. We found no parametric configuration $\{L_n$,$Ca$,$Q\}$ for which the amplitude $\delta F(z)$ of the critical eigenmode showed a significant oscillation of the tapering meniscus, even when $L_n$ was decreased to 7.5. For the sake of illustration, Fig.\ \ref{amplitude2} shows the amplitude $\delta F(z)$ of the critical eigenmode for a flow rate ratio of $Q=0.000125$, smaller than the experimental value $Q_{\textin{min}}=0.00017$ at the jetting-to-dripping transition (Fig.\ \ref{modes}). The amplitude $\delta F$ vanishes in the tapering meniscus, implying that the jet destabilizes while the meniscus remains stable. The real and imaginary parts of $\delta F$ practically coincide. This indicates an in-phase interface oscillation $\delta F(z,t)=|\delta F(z)| \exp(\omega_i t)\cos(\omega_r t+\pi/4)$ corresponding to an oscillation of the outlet flow rate.   

% Conclusion
The failure of the global stability analysis can be attributed to the difference between the experimental and numerical marginally stable base flows. The global stability analysis looks for the parameter conditions for a true steady jetting-to-polydisperse dripping transition. However, this transition does not occur in the experiments because the flow follows a sequence of intermediate monodisperse microdripping states as $Q$ decreases before reaching the polydisperse dripping. This constitutes a fundamental difference with respect to most tip streaming configurations, where a meniscus emitting a steady, long, cylindrical jet destabilizes at $Q=Q_{\textin{min}}$ \citep{CHGM17,RRCHGM21,CRRHM21,LCMH22,PRHGM18}.

\subsection{Convective instability of the jetting mode}
\label{seca2}

\begin{figure}
\begin{center}
\includegraphics[width=1\linewidth]{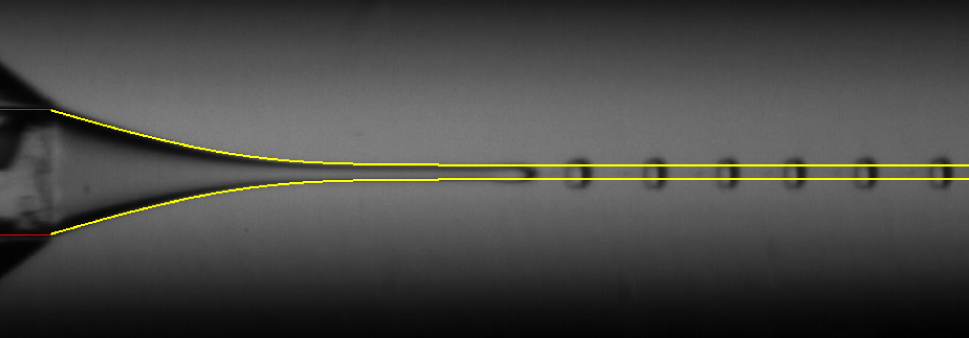}
\end{center}
\caption{Comparison between an experimental image and the numerical base flow for $Ca=0.58$ and $Q=0.00062$.}
\label{exp}
\end{figure}

%  The two cases: Ln=30 and 35
All the simulations in this section were conducted for $Ca=0.58$, and $Q=0.00062$. These parameter conditions correspond to a convectively unstable flow, in which the tapering meniscus remains stable and the emitted jet breaks up within the analyzed fluid domain (Fig.\ \ref{modes}). Figure \ref{exp} shows the remarkable agreement between the numerical steady jetting base flow and the experiment. We considered two values of the outer capillary length, $L_n=15$ and 20, to analyze the validity of the global stability analysis.

\begin{figure}
\begin{center}
\includegraphics[width=7.5cm]{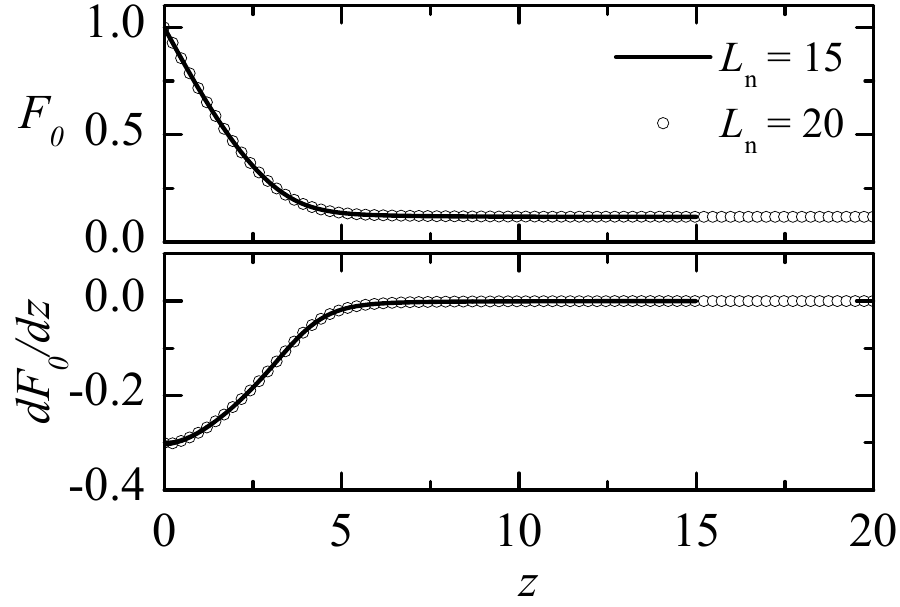}
\end{center}
\caption{Basic flow interface shape $F_0(z)$ (a) and the axial derivative $dF_0/dz$ (b) for $L_n=15$ (symbols) and 20 (solid line).}
\label{radii}
\end{figure}

\begin{figure}
\begin{center}
\includegraphics[width=0.75\linewidth]{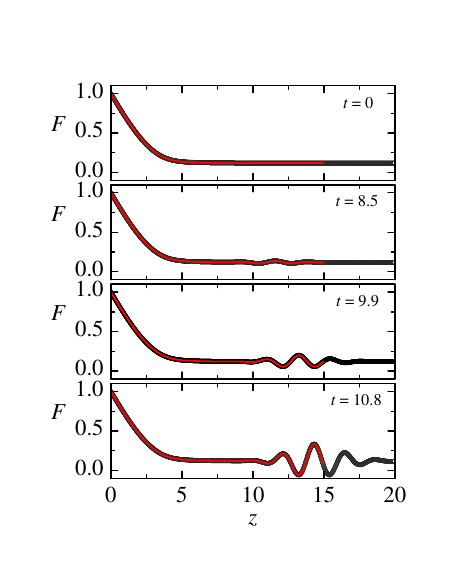}
\end{center}
\caption{Interface location at different instants obtained from the transient simulation for $z_0=1$ and $L_n=15$ (line) and 20 (symbols).}
\label{evolution}
\end{figure}

% Base flow
First, we focus on the results for $L_n=20$ (the symbols in  Figs.\ \ref{radii} and \ref{evolution}). Figure \ref{radii} shows the basic flow interface shape, $F_0(z)$, and the axial derivative $dF_0/dz$. The jet adopts a quasi-cylindrical shape next to the outlet section. We verified that the velocity profile in that section is practically that given by the Hagen-Poiseuille parabolic law for a compound jet \citep{RCMH24}:
\begin{equation}
\label{ej}
R_j=R\left(\frac{Q}{1+Q+\sqrt{1+Q\mu}}\right)^{1/2},
\end{equation}
%\begin{equation}
%\label{K}
%K=-8\, Ca\, (1-R^{-2})\, \mu^{-2} \left[2+\mu(Q+\mu-2)+2(\mu-1)\sqrt{1+Q\mu}\right],
%\end{equation}
\begin{equation}
\label{vs0}
v_s=2\, Ca\, (1-R^{-2})\, \mu^{-2}\left(\mu-1+\sqrt{1+Q\mu}\right),
\end{equation}
where $R_j$ and $v_s$ are the jet radius and surface velocity. Their simulation values are 0.1171 and 0.0312, respectively, while the corresponding analytical predictions (\ref{ej})--(\ref{vs0}) are 0.1176 and 0.0319. We conclude that prescribing the outflow boundary condition (\ref{out}) on the outlet section is fully justified. In other words, one does not expect small variations of $L_n$ to affect the numerical results.

\begin{figure}
\begin{center}
\includegraphics[width=7cm]{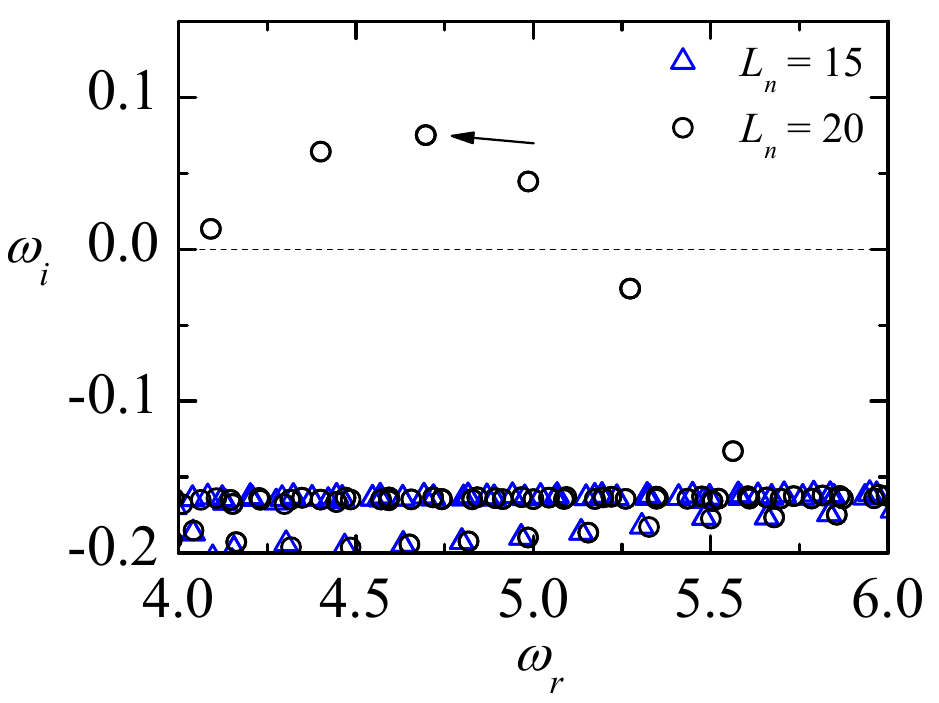}
\end{center}
\caption{Eigenvalues with $4\leq \omega_r\leq 6$ and $\omega_i\geq -0.2$ for $L_n=15$ (triangles) and $L_n=20$ (circles). The black arrow indicates the eigenvalue corresponding to the dominant growing eigenmode.}
\label{eigen}
\end{figure}

\begin{figure}
\begin{center}
\includegraphics[width=7.5cm]{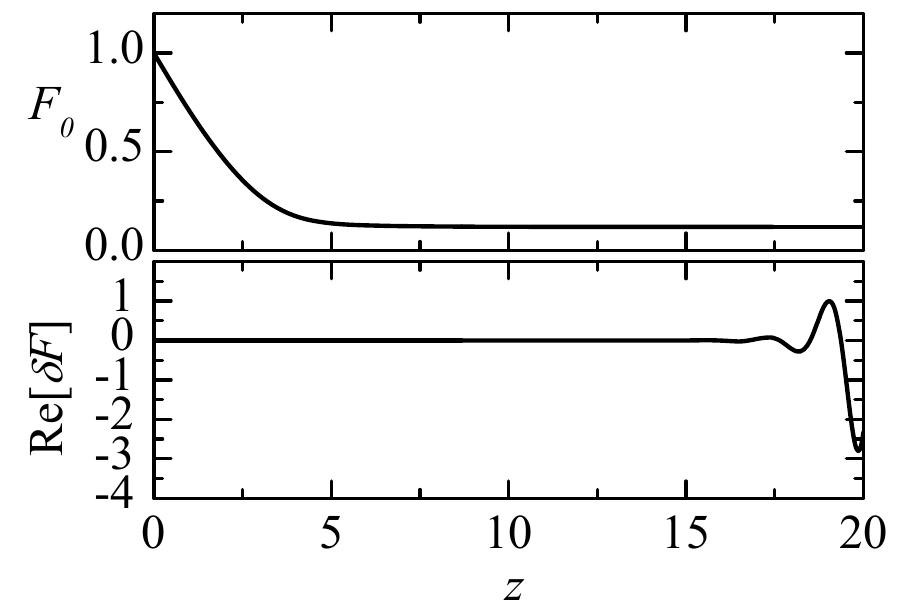}
\end{center}
\caption{Real part of the interface deformation amplitude, Re[$\delta F$], corresponding to the dominant mode for $L_n=20$. The upper graph shows the base flow interface shape.}
\label{amplitude}
\end{figure}

% Eigenvalues and critical eigenmode
The linear stability analysis shows the existence of unstable oscillatory eigenmodes (Fig.\ \ref{eigen}). The amplitude $\delta F$ of the dominant mode indicates that the interface remains practically unaltered for $z\lesssim 15$ (Fig.\ \ref{amplitude}). 

%  Initial perturbation
We conducted a transient simulation of the linearized equations to analyze the interface evolution when a small-amplitude initial perturbation is introduced, as explained in Sec.\ \ref{sec2}. Figure \ref{evolution} shows the interface location at different instants. The interface deformation grows within the interval $11\lesssim z\lesssim 18$ due to the linear superposition of the modes triggered by the perturbation. These results are inconsistent with the linear stability analysis, which predicts that the interface remains practically unaltered for $z\lesssim 15$ (Fig.\ \ref{amplitude}). We conclude that the asymptotic linear stability analysis fails to describe the growth of the perturbation, even though the unstable character of the flow is correctly predicted. As shown below, the success of this prediction is accidental because it relies on spurious growing eigenmodes whose existence depends on the axial length $L_n$ of the numerical domain.

% Case Ln=30
To demonstrate the above statement, we focus on the results obtained for $L_n=15$ indicated by the lines in Figs.\ \ref{radii} and \ref{evolution}. The base flow is practically the same as that for $L_n=20$ (Fig.\ \ref{radii}), confirming that the numerical domain is large enough to ensure the validity of the outflow boundary condition (\ref{out}). The transient numerical simulation shows practically the same interface deformation growth for $L_n=15$ and 20 (Fig.\ \ref{evolution}). However, the unstable eigenmodes disappear from the spectrum for $L_n=15$ (Fig.\ \ref{amplitude}). This occurs because shortening the numerical domain prevents the spatial development of those modes. The linear stability analysis fails to predict the unstable character of the base flow for $L_n=15$. As anticipated, the instability predicted for $L_n=20$ must be attributed to spurious eigenmodes.

\subsection{Superposition of stable eigenmodes}
\label{seca3}

\begin{figure}
\begin{center}
\includegraphics[width=1\linewidth]{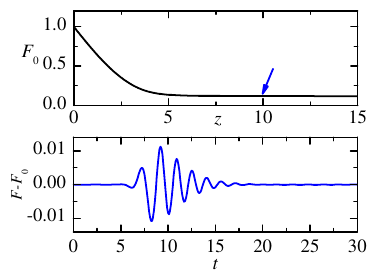}
\end{center}
\caption{Interface deformation $F-F_0$ at $z=10$. The results were obtained from the transient simulation for $L_n=15$ and $z_0=1$. The upper graph shows the base flow shape and location where the interface deformation is shown.}
\label{evolution2}
\end{figure}

%  Superposition of stable eigenmodes
We have concluded that the convective instability of the coflow configuration cannot be explained by the asymptotic linear stability analysis of the steady base flow but by the short-term growth of the perturbation energy due to the superposition of stable eigenmodes. To illustrate this idea, Fig.\ \ref{evolution2} shows the evolution of the interface deformation $F-F_0$ at $z=10$. The fast Fourier transform of $F-F_0$ peaks at around $\omega_r=4.4$, indicating that the superposition of several eigenmodes with frequencies around that value causes the short-term increase in $F-F_0$. These eigenmodes are excited by the initial perturbation and have a sufficiently large amplitude $|\delta F|$ at the selected position $z=10$. 

\begin{figure}
\begin{center}
\includegraphics[width=0.9\linewidth]{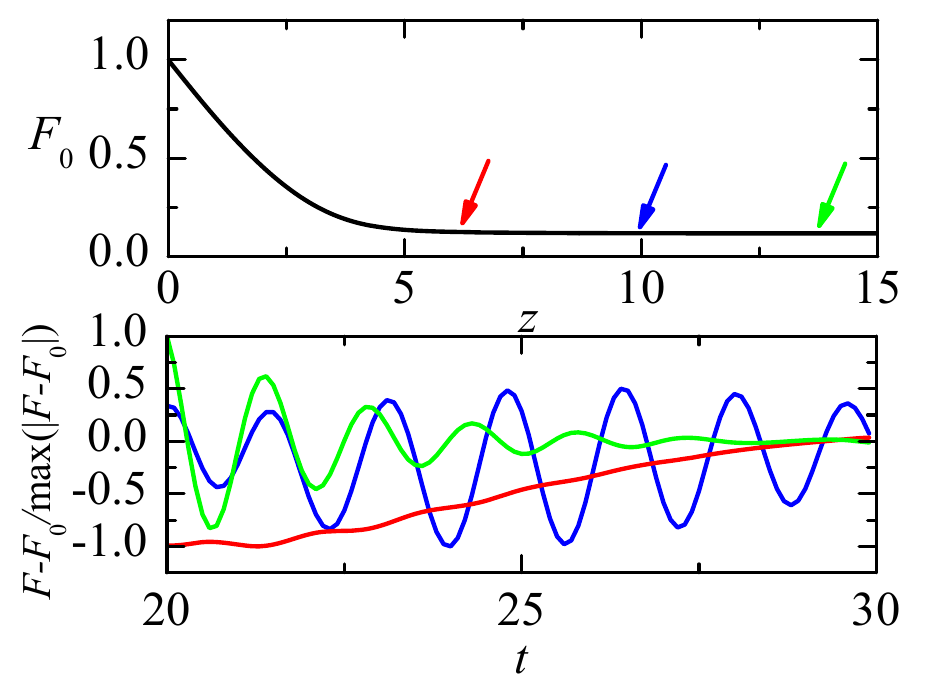}
\end{center}
\caption{Interface deformation $F-F_0$ for $z=6.2$ (red line), 10 (blue line), and 13.8 (green line). The values of $F-F_0$ were normalized by dividing them by the corresponding maximum value. The results were obtained from the transient simulation for $L_n=15$ and $z_0=1$. The upper graph shows the base flow interface shape and the locations where the interface deformation is shown.}
\label{asym}
\end{figure}

% Long-term response
Now, we explore the interface deformation for large $t$. For a sufficiently large initial perturbation, this asymptotic behavior is not observed experimentally because the jet breaks up before. For a sufficiently small initial perturbation, the flow evolution in this regime is expected to be governed by a single (dominant) eigenmode. In this case, the interface deformation would obey the function 
\begin{equation}
F(z,t)-F_0(z)=a(z)\, e^{\omega_i t}\, \cos(\omega_r t+\phi(z))
\end{equation}
with $\omega_r$ and $\omega_i$ giving by the corresponding eigenvalue. We have verified that this is not the case in our simulation (Fig.\ \ref{asym}), suggesting that several eigenmodes coexist during the perturbation damping. This may be anticipated from the spectrum of eigenvalues shown in Fig.\ \ref{eigen}, which shows several eigenvalues with similar frequencies and damping rates. 

\begin{figure}
\begin{center}
\includegraphics[width=0.9\linewidth]{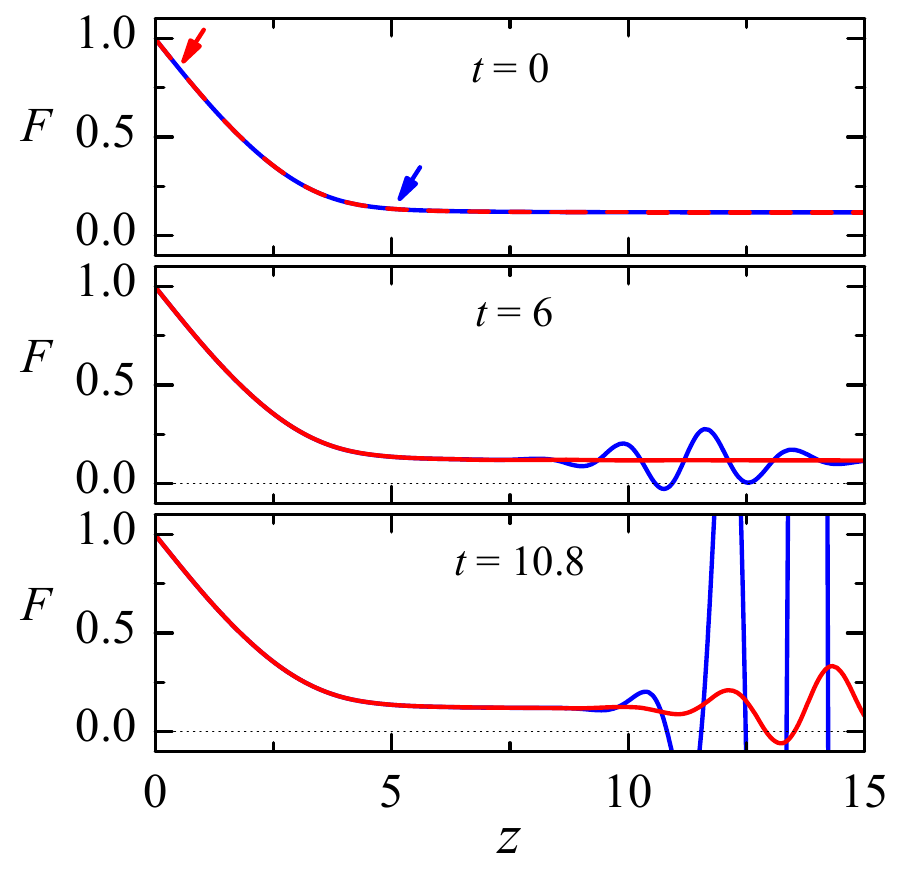}
\end{center}
\caption{Interface location at different instants for $z_0=1$ (red lines) and 5 (blue lines). The results were obtained from the transient simulation for $L_n=15$.}
\label{evolution4}
\end{figure}

% Experimental breakup length
A natural question is whether the transient simulations can be used to determine the jet breakup length. Our results show that this length is not an intrinsic property of the base flow but depends on the perturbation triggering the instability. Figure \ref{evolution4} compares the interface evolution when the initial perturbation is introduced at $z_0=1$ and 5. The base flow is more ``sensitive" to perturbations when they are located in the meniscus-jet transition. For this reason, the interface breakup condition $F=0$ is reached earlier for $z=5$. The breakup point is $z=13.2$ and 10.7 for $z_0=1$ and 5, respectively. This dependency of the breakup point on the initial perturbation explains why the experimental jet breakup length is significantly shorter than that obtained in the simulations (Fig.\ \ref{evolution2}).

\begin{figure}
\begin{center}
\includegraphics[width=0.75\linewidth]{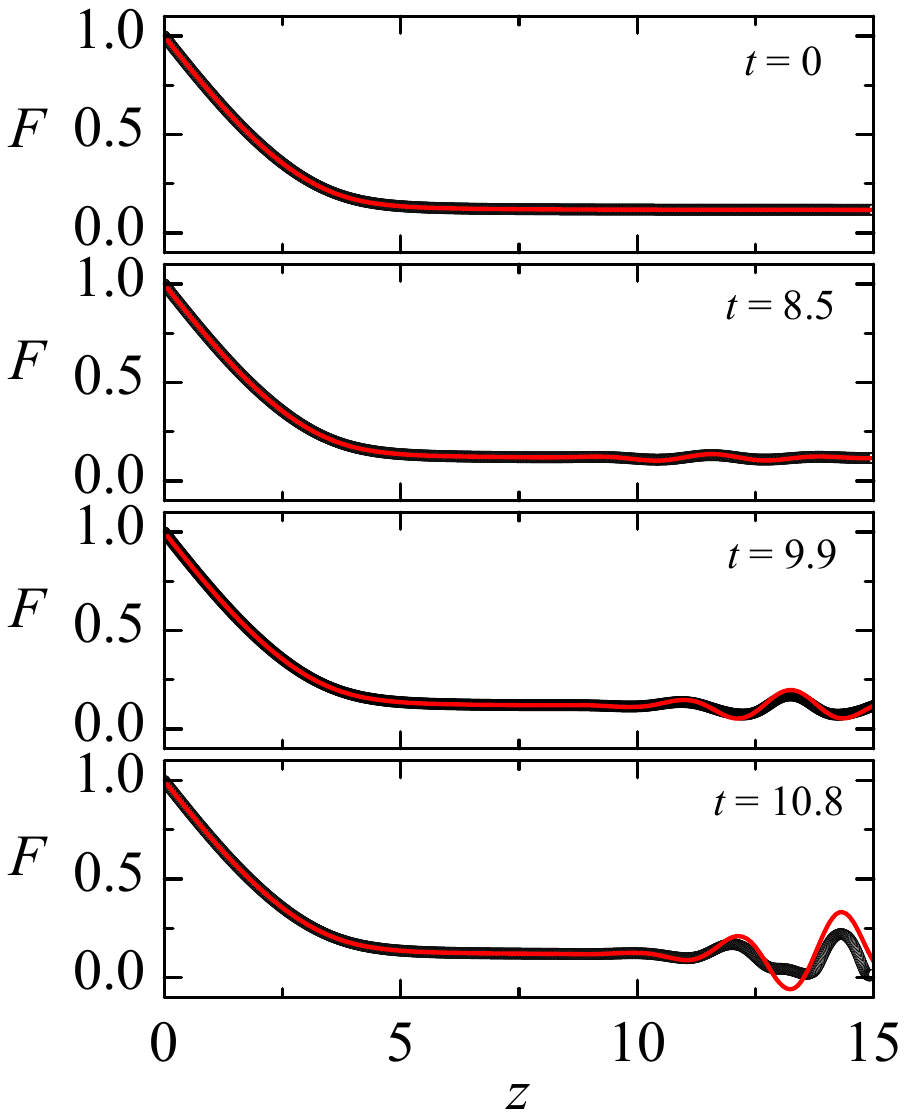}
\end{center}
\caption{Interface location at different instants. The line and symbols correspond to the linear and nonlinear results. The results were obtained for $L_n=15$ and $z_0=1$.}
\label{evolution3}
\end{figure}

% Nonlinear effects
Finally, Fig.\ \ref{evolution3} compares the interface location obtained by integrating the linearized and nonlinear equations. Nonlinear effects become noticeable only close to the interface breakup instant. The growth of the linear perturbation produces a neck and two bulges on the two sides of the neck. The liquid moves from the neck towards the bulges, driven by the capillary pressure, accelerating in the direction of motion. In the nonlinear phase of the interface breakup, the acceleration takes its maximum value in a section located between the neck and the bulge, which makes the central neck symmetrically split into two. Each neck migrates towards the closest bulge until it pinches the interface \citep{CS97}. The pinching point slightly moves downstream due to nonlinear effects.

\section{Conclusions}
\label{sec5}

% Absolute instability
Our results reveal that the global stability analysis of the steady jetting mode fails to predict the flow's absolute instability (jetting-to-polydisperse dripping transition) in a microfluidic coflow device. This occurs because the true base flow at the instability transition corresponds to monodisperse microdripping, which significantly differs from the steady jetting assumed in the stability analysis. In other words, as the inner flow rate decreases, the flow undergoes an intermediate jetting-to-monodisperse microdripping transition, invalidating the analysis of the jetting-to-polydisperse dripping transition. That intermediate transition has not been observed in other tip streaming flows \citep{CHGM17,RRCHGM21,CRRHM21,LCMH22,PRHGM18}. In these flows, the tapering meniscus becomes unstable under parameter conditions for which the jet remains stable, probably due to the focusing effect not present in a coflow device.

% Convective instability
We have shown that the convective instability giving rise to the jetting-to-monodisperse microdripping transition in the coflow configuration cannot be described using the global linear stability analysis either. The growth/decay of the critical eigenmode does not determine the flow stability character owing to the non-normal character of the linearized Navier-Stokes operator. The superposition of the linear modes triggered by an initial perturbation renders the flow unstable even though those modes decay over time. 

The above conclusion has relevant consequences. Specifically, the results derived from the classical linear stability analysis may not apply to the coflow configuration considered in this work. For instance, the jet's absolute instability transition calculated from Brigg's pinch condition \citep{B64,HM90a} may constitute a necessary but insufficient condition for the jet stability in this microfluidic device. This may apply to other similar configurations and explain the discrepancies between this theory and experiments (see Fig.\ 9.6 of Ref.\ \citep{M24}).  

% Jet breakup
The jet breakup length is usually estimated from the classical temporal linear stability analysis of the jet \citep{GG09a}. In this case, one assumes that the dominant temporal mode is responsible for the breakup. This mode is supposed to be triggered by a perturbation next to the jet inception region and convected by the jet. The breakup length scales as the jet velocity times the inverse of the dominant mode growth rate \citep{IYXS18}. The initial perturbation amplitude is a free parameter, which implies that the prefactor of the jet length scaling cannot be predicted. Our results invalidate this approximation in the coflowing configuration because the jet breakup results from the constructive interference of several decaying modes.

\vspace{1cm}
{\bf Declaration of Interests}. The authors report no conflict of interest.

\vspace{1cm}
{\bf Acknowledgement.} This research has been supported by the Spanish Ministry of Economy, Industry, and Competitiveness under Grants PID2019-108278RB, PID2022-140951OB-C21 and PID2022-140951OB-C22/AEI/10.13039/501100011033/FEDER,UE.

% Bibliography
%\bibliography{central} \end{document}


\end{document}